# Electro-opto-mechanical radio-frequency oscillator driven by guided acoustic waves in standard single-mode fiber

*Yosef London,[1] Hilel Hagai Diamandi,[1,a] and Avi Zadok[1,b]*

[1]Faculty of Engineering and Institute for Nano-Technology and Advanced Materials, Bar-Ilan University, Ramat-Gan 5290002, Israel

**Abstract**

An opto-electronic radio-frequency oscillator that is based on forward scattering by the guided acoustic modes of a standard single-mode optical fiber is proposed and demonstrated. An optical pump wave is used to stimulate narrowband, resonant guided acoustic modes, which introduce phase modulation to a co-propagating optical probe wave. The phase modulation is converted to an intensity signal at the output of a Sagnac interferometer loop. The intensity waveform is detected, amplified and driven back to modulate the optical pump. Oscillations are achieved at a frequency of 319 MHz, which matches the resonance of the acoustic mode that provides the largest phase modulation of the probe wave. Oscillations at the frequencies of competing acoustic modes are suppressed by at least 40 dB. The linewidth of the acoustic resonance is sufficiently narrow to provide oscillations at a single longitudinal mode of the hybrid cavity. Competing longitudinal modes are suppressed by at least 38 dB as well. Unlike other opto-electronic oscillators, no radio-frequency filtering is required within the hybrid cavity. The frequency of oscillations is entirely determined by the fiber opto-mechanics.

**Main Text**

Opto-electronic oscillators (OEOs) are sources of radio-frequency (RF) tones, which combine optical and electrical paths within a closed-loop, hybrid cavity [1-4]. In a typical arrangement light from a laser diode source is amplitude-modulated by an RF waveform, and propagates along a section of fiber. The light wave is detected at the output of the fiber, and the recovered RF signal is fed back to modulate the optical input. When the feedback gain is sufficiently high, stable self-sustained RF oscillations may be achieved. OEOs can provide RF tones with extremely low phase noise [3], and they are pursued for applications in coherent communication, radars and precision metrology [4]. OEO cavities are typically long, and support a large number of longitudinal modes that are separated by a comparatively narrow free spectral range. The selection of the specific frequency of oscillations is often aided by a narrow inline RF filter.

---

[a] Y. London and H. H. Diamandi contributed equally to this work.
[b] Author to whom all correspondence should be addressed. Electronic mail: Avinoam.Zadok@biu.ac.il

Various photonic devices exhibit mechanical resonances at radio and microwave frequencies. The interactions between guided light waves and these mechanical modes may give rise to microwaves generation. In one example, the coupling between whispering-gallery optical and mechanical modes in a silica micro-sphere led to oscillations at 11 GHz [5]. Backward stimulated Brillouin scattering (SBS) processes in wedge-shaped silica resonators [6,7], chalcogenide glass waveguides [8], and standard optical fibers [9,10] were also used in the generation of microwaves up to 40 GHz frequencies.

Optical fibers also support guided acoustic modes, at frequencies between hundreds of MHz and several GHz [11-21]. The acoustic modes may be optically stimulated through electro-strictive forces, and can scatter guided light waves via photo-elasticity [12-13]. The phenomena are referred to as guided acoustic waves Brillouin scattering (GAWBS) [12], or as forward stimulated Raman-like, inter-polarization or inter-modal scattering [15]. Photonic crystal fibers (PCFs) [14-17] and micro-structured fibers [18-20] support very strong interactions between guided acoustic and optical modes. The effects were analyzed and applied in a series of papers by Russell and co-workers [14-20]. Remarkable demonstrations include the generation of radio-frequency combs, without feedback, over very short fiber segments [19,20]; and the ultra-stable mode-locking of soliton fiber lasers, which led to long-term bit storage [17]. These works, however, rely on specialty fibers, and do not implement an OEO configuration.

In this work we demonstrate stable RF oscillations that are driven by GAWBS in a standard single-mode fiber (SMF). An optical pump wave is used to stimulate acoustic vibrations. These vibrations, in turn, induce phase modulation to a co-propagating optical probe wave of a different wavelength. The phase modulation spectrum consists of a series of narrowband resonances, each corresponding to a different acoustic mode. The magnitude of GAWBS in SMF is orders of magnitude weaker than in PCFs and in micro-structured fibers, and it is not sufficient to introduce self-oscillations. To work around this limitation, feedback is provided in the form of a hybrid electro-opto-mechanical cavity. A Sagnac loop configuration, which was initially proposed towards the characterization of GAWBS [14,18], is used to convert the probe phase modulation to an intensity signal. The detected waveform is then amplified and fed back to modulate the pump wave. With sufficient optical pump power and RF amplification, the cavity reaches stable oscillations. Unlike most OEO realizations, the frequency of operation is entirely determined by the fiber opto-mechanics. No electrical RF filter is used.

The principle is experimentally demonstrated using a 200 meters long section of SMF. Stable oscillations are obtained at 319 MHz, the frequency of the guided acoustic mode for which phase modulation of the probe wave is the most efficient. Scattering due to competing, weaker acoustic modes does not match the hybrid cavity losses, and oscillations at the corresponding frequencies are strongly suppressed. Furthermore, the linewidth of the acoustic resonance is narrow enough to support oscillations at a single longitudinal mode of the hybrid cavity. Oscillations at the frequencies of adjacent longitudinal modes are strongly suppressed as well.

Optical fibers support several classes of guided acoustic modes. Optical stimulation and scattering in SMF is the most efficient for radial modes, denoted herein as $R_{0m}$ where $m$ is an integer. Each mode is characterized by a cut-off frequency $\Omega_m$ and a linewidth $\Gamma_m$ [12,13]. The material displacement in GAWBS by $R_{0m}$ modes is purely radial, and the associated strain field is entirely transverse. The normalized displacement profile is given by [12,13]:

$$u_r^{(m)}(r) = \frac{J_1\left[(\Omega_m r)/v_d\right]}{\sqrt{2\pi \int_0^a \left\{J_1\left[(\Omega_m r)/v_d\right]\right\}^2 r \, dr}} . \tag{1}$$

Here $a$ is the cladding radius, $v_d$ is the speed of longitudinal sound waves in silica, and $0 \leq r \leq a$ denotes the radial coordinate.

The acoustic modes are stimulated by an optical pump wave of instantaneous power $P(t)$, where $t$ stands for time. The Fourier transform of $P(t)$ is noted by $\tilde{P}(\Omega)$, with $\Omega$ an RF variable. The pump wave propagates along a fiber of length $L$. The transverse strain profile of mode $R_{0m}$ induces photo-elastic perturbations to the effective index of a co-propagating optical probe wave, which lead to the modulation of its phase [12-13,16,21]:

$$\delta\tilde{\phi}_{OM}(\Omega) = \sum_m \frac{k_0}{4n^2 c \rho_0} \frac{Q_{ES}^{(m)} Q_{PE}^{(m)}}{\Gamma_m \Omega_m} L \frac{1}{j - 2(\Omega - \Omega_m)/\Gamma_m} \tilde{P}(\Omega) . \tag{2}$$

The phase modulation spectrum $\left|\delta\tilde{\phi}_{OM}(\Omega)\right|^2$ consists of a series of narrowband resonances at frequencies $\{\Omega_m\}$ [12-13,16,21]. It is independent of the states of polarization (SOPs) of the pump and probe waves. In Eq. (2), $\rho_0$ and $n$ are the density and refractive index of silica respectively, $k_0$ is the vacuum wavenumber of the probe wave, and $c$ is the speed of light in vacuum. The electro-strictive overlap integral $Q_{ES}^{(m)}$ determines the efficiency of stimulation of mode $R_{0m}$ [13]:

$$Q_{ES}^{(m)} \equiv (a_1 + 4a_2) \cdot 2\pi \int_0^a E_T(r) \frac{dE_T(r)}{dr} u_r^{(m)}(r) r dr. \qquad (3)$$

Here $a_1$ = 0.66 and $a_2$ = -1.2 are photo-elastic parameters of silica, and $E_T(r)$ is the normalized transverse profile of the electric field of the optical mode. A Gaussian profile, defined by its mode field diameter (MFD), is assumed in this work. Lastly, the photo-elastic overlap integral $Q_{PE}^{(m)}$ describes the modification of the probe's effective index by mode $R_{0m}$ [13]:

$$Q_{PE}^{(m)} \equiv \left(\frac{a_1}{2} + a_2\right) \cdot 2\pi \int_0^a \left[\frac{du_r^{(m)}(r)}{dr} + \frac{u_r^{(m)}(r)}{r}\right] |E_T(r)|^2 r dr. \qquad (4)$$

The instantaneous phase modulation of the probe wave $\delta\phi_{OM}(t)$ is obtained by the inverse Fourier transform of Eq. (2). The magnitude of probe wave phase modulation scales with the pump power and with the fiber length. This dependence is analogous to that of phase modulation due to the Kerr effect. It is therefore instructive to define a nonlinear coefficient, equivalent to the γ parameter used to quantify the Kerr effect in fiber, in units of [W×km]$^{-1}$ [16]:

$$\gamma_{OM}^{(m)} \equiv \frac{k_0}{4n^2 c \rho_0} \frac{Q_{ES}^{(m)} Q_{PE}^{(m)}}{\Gamma_m \Omega_m}. \qquad (5)$$

At the acoustic resonance frequencies, we obtain [16]:

$$\left|\delta\tilde{\phi}_{OM}^{(m)}(\Omega_m)\right|^2 = \left|\gamma_{OM}^{(m)}\right|^2 L^2 \left|\tilde{P}(\Omega_m)\right|^2. \qquad (6)$$

Opto-mechanical phase modulation was measured experimentally in a 200 meters-long SMF section and in a 1 km-long commercial highly nonlinear fiber (HNLF). Both fibers were subsequently used in demonstrations of electro-opto-mechanical oscillations. (For details of the measurement setup and procedures, see [14,18,21]). Figure 1(a) shows the measured normalized PSD of the input pump power $|\tilde{P}(\Omega)|^2$, and the measured normalized PSD of the probe wave phase modulation $|\delta\tilde{\phi}_{OM}(\Omega)|^2$ in SMF. The relative magnitudes of the nonlinear coefficients $|\gamma_{OM}^{(m)}|$ were estimated based on the ratio between the measured $|\delta\tilde{\phi}_{OM}(\Omega_m)|^2$ and $|\tilde{P}(\Omega_m)|^2$, as in Eq. (6), and calculated analytically using the model of Eq. (5). The measured and calculated normalized coefficients for SMF (MFD = 9.6 μm) and for HNLF (MFD = 4.2 μm) are shown in Fig. 1(b) and Fig. 1(c), respectively. Agreement between measurement and analysis is very good, for both fibers. The strongest effect in SMF was observed for mode $R_{0,7}$ at a frequency of 319 MHz. The measured modal linewidth $\Gamma_7$ was 2π·4.8 MHz, and analysis suggested a value of $|\gamma_{OM}^{(7)}|$ = 5 [W×km]$^{-1}$. The strongest GAWBS in the HNLF takes place through modes $R_{0,14}$ and $R_{0,15}$, with $\Omega_{14}/(2\pi)$ = 651 MHz, $\Omega_{15}/(2\pi)$ = 699 MHz, $\Gamma_{14} \approx \Gamma_{15}$ = 2π·8 MHz, and calculated nonlinear coefficients $|\gamma_{OM}^{(14)}| \approx |\gamma_{OM}^{(15)}|$ = 16.6 [W×km]$^{-1}$.

Figure 2 shows the experimental setup for an OEO that is driven by guided acoustic waves. Light from a first laser diode at 1553 nm wavelength passed through an electro-optic amplitude modulator (EOM, $V_\pi$ = 5 V), was then amplified by an erbium-doped fiber amplifier (EDFA) to an average power of $P_p$ = 32 dBm, and launched into one end of the 200 m-long SMF as a pump wave. The fiber was placed within a Sagnac loop configuration [14,18,21]. Light from a second laser diode at 1550 nm wavelength was used as a probe wave. The probe wave propagated along the fiber in both directions, whereas the pump wave propagated in the clockwise (CW) direction only. The Sagnac loop eliminated variations in the output probe due to environmental, reciprocal phase drifts.

Due to the wave-vector matching characteristics of GAWBS, phase modulation to the probe wave is non-reciprocal [12-14,18,21]: modulation to the CW probe wave accumulated over the length $L$, whereas the overall modulation of the counter-clockwise (CCW) probe wave was negligible. Non-reciprocal phase modulation was converted to an intensity waveform upon detection of the probe wave at the output of the Sagnac loop. Polarization controllers (PCs) were used to adjust the SOP of the input probe wave, and those of the CW and CCW propagating probe waves within the loop. PCs were aligned in attempt to maximize the RF power of oscillations and the suppression of side-modes. Optimal loop bias was more readily achieved with the manual adjustment of two PCs rather than one. Optical bandpass filters were used to block the pump wave, and undesired backwards SBS of the pump wave, from reaching the detector. When $\delta\phi_{OM}(t) \ll \pi$, the frequency-domain output voltage of the detector may be expressed as:

$$\delta\tilde{V}(\Omega) \approx \alpha R P_s^{bias} \delta\tilde{\phi}_{OM}(\Omega) \ . \tag{7}$$

Here $P_s^{bias}$ = 1.5 mW is the probe wave power at the loop output in the absence of pump, and with the loop biased at quadrature. Also in Eq. (7), $R$ = 27 V/W is the responsivity of the photo-detector and its associated circuitry, and $\alpha \leq 1$ is a scaling factor that depends on the alignment of SOPs. The magnitude $|\delta\tilde{V}(\Omega)|$ is the largest at the acoustic resonance frequencies $\{\Omega_m\}$.

The detector output voltage was amplified by an RF amplifier with a small-signal voltage gain $G_{RF}$ of 25, and used to drive the pump wave EOM. In this manner, an opto-electronic feedback loop was closed between the pump and probe waves, with the intermediation of guided acoustic waves in the fiber. Sufficient feedback could drive the initial RF noise of the electrical amplifier and thermal vibrations of the acoustic modes into oscillations. A 10% coupler and a second photo-detector were used for monitoring the oscillating waveform.

Figure 3, panels (a) through (d) show measurements of the OEO output PSD $|\delta\tilde{V}(\Omega)|^2$, taken at several frequency scales. Stable oscillations were obtained at a frequency of 319.06 MHz,

which closely matches $\Omega_7/(2\pi)$. Secondary peaks observed in panel (a) are at the resonance frequencies of several other acoustic modes (see also Fig. 1(a)), however the RF power in those was suppressed by at least 40 dB with respect to the primary frequency. View of the output PSD on a finer scale (panel (b)) reveals multiple frequencies within the linewidth $\Gamma_7$ of the dominant acoustic mode. These are due to multiple longitudinal modes of the hybrid cavity. They are separated by a free spectral range of 654 kHz, which corresponds to the overall round-trip time delay in the cavity. Note that this delay is longer than that of the fiber loop alone, since the cavity also consists of fiber amplifiers and components in the pump and signal paths. All longitudinal side-modes within the acoustic linewidth are suppressed by at least 38 dB with respect to the main one. Stable, single-mode OEO operation was therefore achieved, based on GAWBS alone.

The full width at half maximum linewidth of the output spectrum was 300 Hz. The relative phase noise of the oscillator was measured as -94 dBc/Hz at an offset of 10 kHz. The phase noise of the RF spectrum analyzer available to us is specified as -95 dBc/Hz at the same offset. Hence the actual phase noise of the OEO might have been better. The power levels at high-order harmonics of $\Omega_7/(2\pi)$ were 26 dB weaker than that of the fundamental frequency (panel (d)). Measurements were also performed with the SMF section replaced by the 1 km-long HNLF (pump power $P_p = 25$ dBm, output probe power $P_s^{bias} = 0.5$ mW, RF voltage gain $G_{RF} = 1000$). Oscillations were obtained at a frequency of 648.7 MHz, which closely matches $\Omega_{14}/(2\pi)$ (panel (e)). Here too, oscillations were locked to a single acoustic resonance and to a single longitudinal mode within that resonance.

The SMF-based oscillator was locked to $\Omega_7/(2\pi)$ in practically all manual adjustments of PCs states. Oscillations at $\Omega_8/(2\pi) = 369$ MHz were observed at very few states, and locking to the frequencies of other acoustic mode could not be achieved. Adjustments of the PCs also induced switching among adjacent longitudinal modes within $\Gamma_7$. Tuning was not predictable, and the mechanism underlying polarization-induced switching was not investigated. Potential explanations include nonlinear polarization rotation, arbitrary interference with spontaneous emission of optical amplifiers in the probe and pump paths, or modifications to the RF phase delays acquired in a hybrid cavity round trip. The HNLF-based oscillator could be switched among the frequencies of several acoustic resonances, between $\Omega_{10}/(2\pi)$ (463 MHz) and $\Omega_{15}/(2\pi)$ (699 MHz), due to the smaller differences among the nonlinear coefficients $\left|\gamma_{OM}^{(m)}\right|$ of adjacent modes in that fiber (see Fig. 1(c)). Temperature variations change $\{\Omega_m\}$ by 93 ppm per °K [22], and modify the group delay in the fiber loop by 7.5 ppm per °K [23]. Both mechanisms may lead to switching among longitudinal modes. Vibration-induced reciprocal phase delays are cancelled at the Sagnac loop output. On the other hand, residual vibrations-induced birefringence may broaden the oscillations linewidths.

In conclusion, electro-opto-mechanical RF oscillations in SMF were proposed and demonstrated experimentally. Oscillations are reached due to forward, intra-modal scattering by guided acoustic modes of the fiber, which couple the optical pump and probe waves. The PSD of scattering by guided acoustic waves consists of a series of resonances, which are sufficiently narrow to restrict the operation of the OEO to a single frequency with no electrical RF filtering. The RF waveform propagates as an electrical signal along part of the hybrid cavity, takes up the form of a modulated optical carrier in another part, and also assumes the form of a mechanical vibration. The results can provide a link between the OEO community, who works in SMF and seldom considers guided acoustic waves, and the opto-mechanics community, who for the most part works in specialty fibers and does not focus on OEOs. Both communities might consider the opto-mechanics of SMF too weak to be employed. The method presented herein works around that difficulty.

This concept has several limitations. Tuning of the oscillations frequency is very limited, opto-mechanical interactions are much weaker than those observed in PCFs or in micro-structured fibers, and the phase noise of the RF output is not as low as those of state-of-the-art OEOs that have been developed over 20 years [2-3,24]. However, the use of GAWBS towards frequency discrimination in OEOs can be regarded as a complementary mechanism, rather than a competing one. Opto-mechanics may be incorporated alongside existing low-phase-noise techniques [24], such the use of dual fiber loops, in attempt to enhance performance even further. The phase noise performance of OEOs is known to improve with the cavity propagation delay [2,3]. Long cavities would be simpler to implement based on SMF, which is readily available to everyone, than in using micro-structured fibers or PCFs. Future work would include a comprehensive noise analysis, and address the incorporation of GAWBS in advanced OEO setups [24].

## Acknowledgements

The authors thank Mr. Eyal Preter of Bar-Ilan University for assistance with data analysis. This work was supported in part by a Starter Grant from the European Research Council (ERC), Grant number H2020-ERC-2015-STG 679228 (L-SID), and by the Israel Science Foundation (ISF), Grant number 1665-14.

[22] Y. Tanaka, K. Ogusu, *IEEE Photon. Technol. Lett*. **10**, 1769 (1998).

[23] A. Ben-Amram, Y. Stern, Y. London, Y. Antman, A. Zadok, *Opt. Express* **23**, 28244 (2015).

[24] E. Levy, K. Okusaga, M. Horowitz, C. R. Menyuk, W. Zhou, G. M. Carter, *Opt. Express* **18**, 21461 (2010).
12

**Figures**

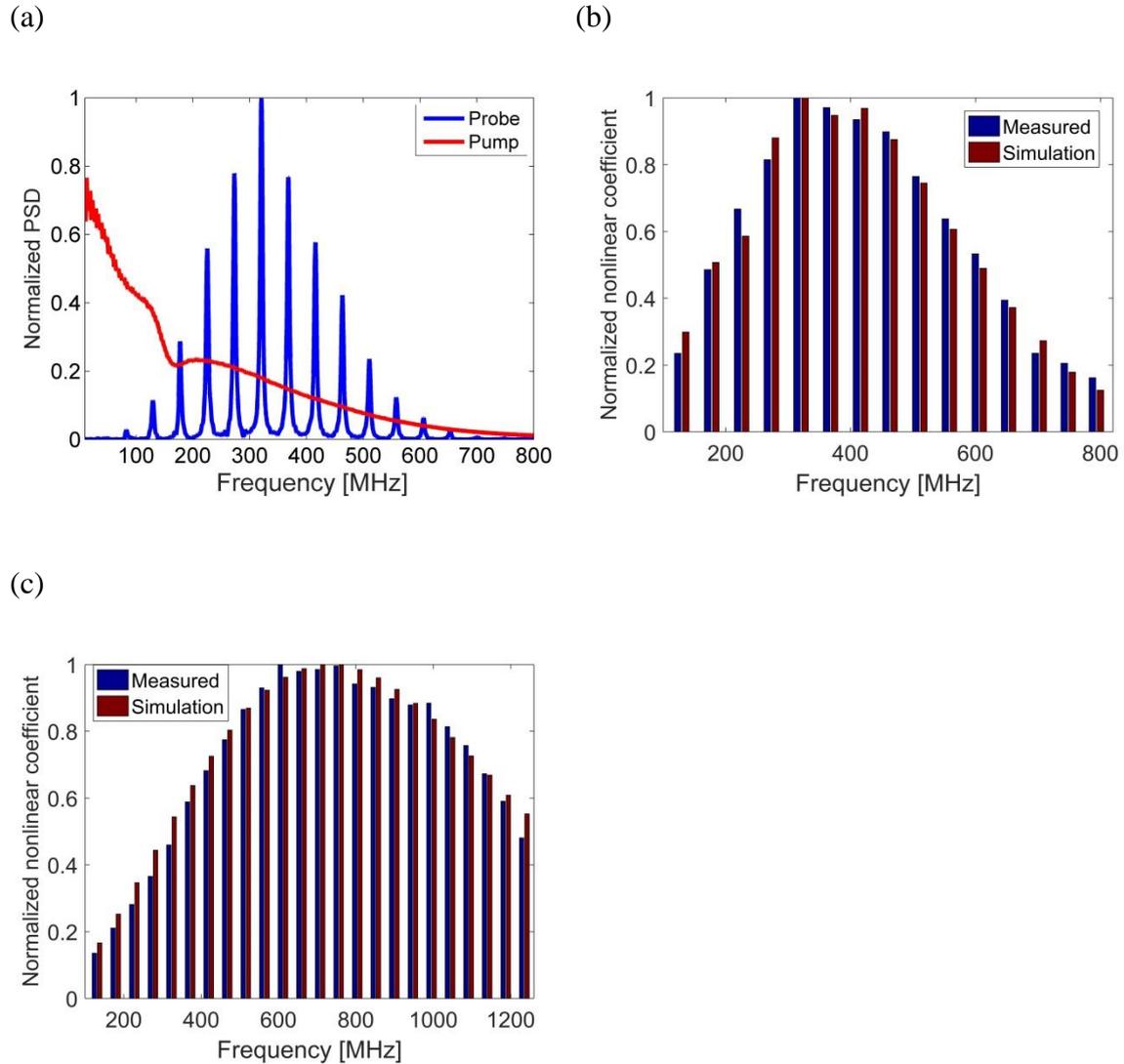

Figure 1. (a) – Measured normalized power spectral density of the input pump wave (red); measured power spectral density of the resulting probe wave phase modulation in SMF, due to guided acoustic waves (blue). (b) – measured (blue) and calculated (red) relative nonlinear coefficients of the probe wave opto-mechanical phase modulation in SMF, as a function of the resonant frequencies of guided acoustic modes. The strongest modulation is driven by mode $R_{0,7}$, at a resonance frequency of 319 MHz. (c) – same as panel (b), with the SMF replaced by a HNLF. The strongest modulation in that fiber is driven by modes $R_{0,14}$ and $R_{0,15}$, at resonance frequencies of 651 MHz and 699 MHz, respectively.

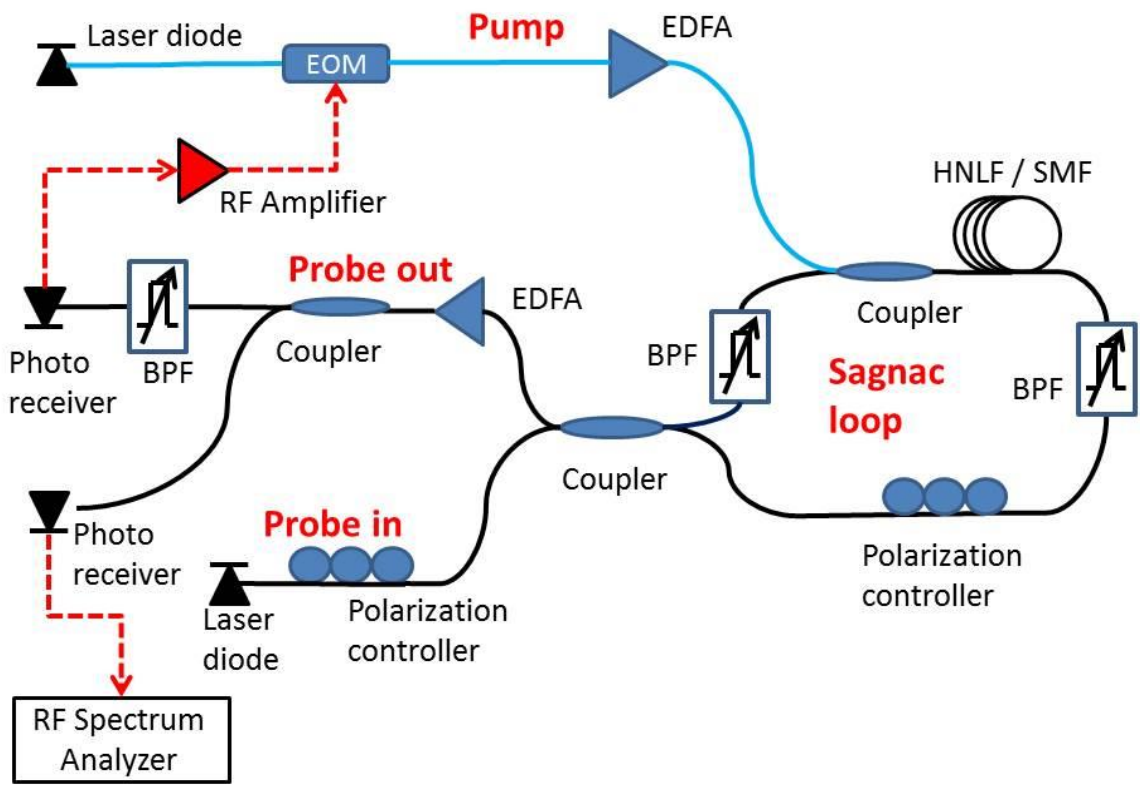

Figure 2. Experimental setup used in the demonstration of an electro-opto-mechanical radio-frequency oscillator over fiber. EDFA: erbium-doped fiber amplifier. BPF: optical bandpass filter. RF: radio frequency. SMF: single-mode fiber. HNLF: highly nonlinear fiber.

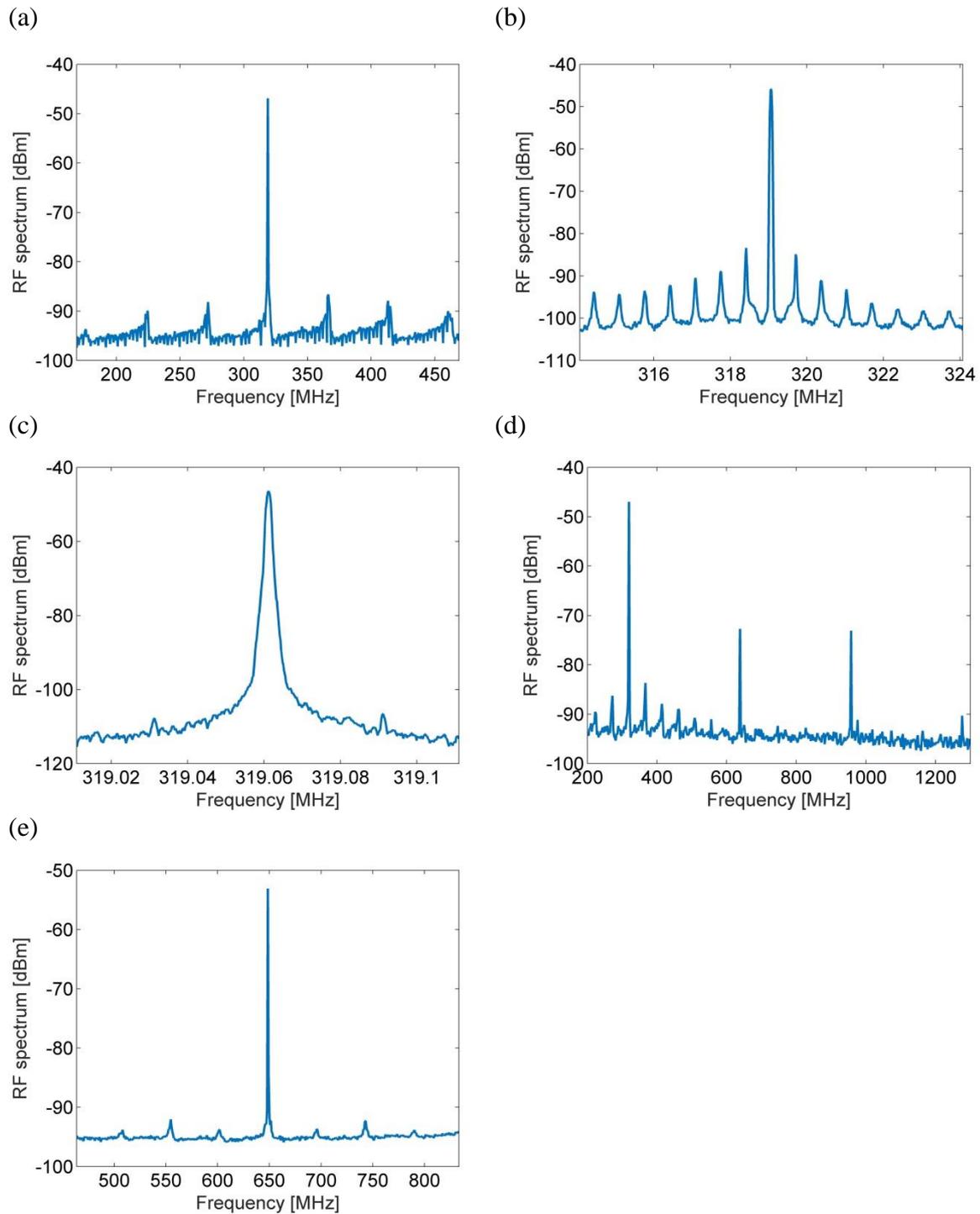

Figure 3. (a) through (d) - Measured power spectral density of the output voltage of an electro-opto-mechanical radio-frequency oscillator, consisted of a 200 m-long SMF. Stable oscillations are obtained at a fundamental frequency of 319.06 MHz. Panels (a) through (d) correspond to measurements at different frequency scales. The peak at 319 MHz in panel (d) corresponds to the fundamental frequency of oscillations. Secondary peaks at 638 MHz, 957 MHz and 1.276 GHz are high-order harmonics. (e) – Measured power spectral density of the output voltage, with the SMF replaced by a 1 km-long section of HNLF. Oscillations are obtained at 648.7 MHz.